\documentclass{article}
\usepackage{ICASSP2026_Paper_Templates/spconf,amsmath,graphicx,hyperref}
\usepackage{amsfonts}
\usepackage{stmaryrd}
\usepackage{graphicx}
\usepackage{fontawesome5}
\usepackage{xcolor}
\usepackage{esvect}
\usepackage{indentfirst}

\title{Sobolev Norms in Neural Embeddings Measure Audio Morphing Regularity}
\name{Théo Chasle Cauchy\textsuperscript{1}, Modan Tailleur\textsuperscript{1}, Barbara Pascal\textsuperscript{1}, Fanny Roche\textsuperscript{2}, Mathieu Lagrange\textsuperscript{1}}
\address{\textsuperscript{1}Nantes Université, École Centrale Nantes, CNRS, LS2N, UMR6004, F-44000 Nantes, France.\\  \textsuperscript{2}Arturia, Montbonnot Saint-Martin, France}
\def\Refer{$\mathsf{LUM}$}
\def\Null{$\mathsf{Null}$}

\def\Corres{$\mathsf{Correspondence}$}
\def\Smooth{$\mathsf{Smoothness}$}
\def\Interm{$\mathsf{Intermediateness}$}

\def\SmoothMF{$\mathsf{Smoothness}$ $\mathsf{MF}$}
\def\SmoothSM{$\mathsf{Smoothness}$ $\mathsf{SM}$}
\def\IntermSM{$\mathsf{Intermediateness}$ $\mathsf{SM}$}
\def\CorresSM{$\mathsf{Correspondence}$ $\mathsf{SM}$}
\def\corres{$\mathsf{Corresp.}$ $\mathsf{SM}$}
\def\smoothMF{$\mathsf{Smooth.}$ $\mathsf{MF}$}
\def\smoothSM{$\mathsf{Smooth.}$ $\mathsf{SM}$}
\def\interm{$\mathsf{Interm.}$ $\mathsf{SM}$}

\def\NUC{$\mathsf{NUC}$}
\def\EQC{$\mathsf{EQC}$}

\begin{document}
\ninept
%

\setlength{\belowdisplayskip}{0pt} \setlength{\belowdisplayshortskip}{0pt}
\setlength{\abovedisplayskip}{0pt} \setlength{\abovedisplayshortskip}{0pt}


\maketitle
\begin{abstract}

Morphing has recently gained renewed interest with the emergence of generative models, particularly in audio and image generation. In musical sound synthesis, morphing can generate intermediate sounds between two targets, helping musicians and sound engineers explore new sounds with interesting perceptual properties. As morphing is inherently defined in perceptual terms, evaluating this task is challenging. In this work, we introduce Sobolev Distances to Ideal Morphing (SDIM), a novel objective metric to quantify the regularity of audio morphing trajectories in perceptually relevant audio embedding spaces. Leveraging a physics-based sound synthesizer, we evaluate the discriminative power of SDIM on controlled morphing trajectories with varying degrees of regularity and compare it with that of existing audio morphing metrics. Results show that, contrary to state-of-the-art metrics, the proposed metric reliably discriminates desirable trajectories from adversarial ones. 




\end{abstract}
\section{Introduction}

Morphing is defined in \cite{slaney1996automatic} as the process of smoothly transforming one entity into another, by gradually interpolating between their dissimilarities while preserving their shared characteristics. The morphing task has recently attracted renewed interest in both image and audio generation, particularly with the emergence of latent space generative models, among which diffusion models, which inherently enable morphing by interpolating their input noise vectors \cite{wang2023interpolating, zhang2024diffmorpher, niu2024soundmorpher, kamath2025morphfader, chu2026mix2morph}. The evaluation of a procedure that performs the morphing task requires assessing whether the smooth evolution of the input parameters along the morphing trajectories between pairs of starting and ending points does translate into a perceptually progressive modification. The perceptual human-based evaluation of morphing is cumbersome as it requires evaluating a large number of morphing trajectories, and might suffer from large variance across subjects. Consequently, the complementary use and design of efficient and objective metrics is of great interest. 

In image processing, several metrics have been proposed to assess different aspects of morphing. The perceptual path length measures the perceptual smoothness of trajectories in the latent space of generative models \cite{karras2019style}. The perceptual image patch similarity has also been leveraged to evaluate morphing quality by summing the distances between consecutive generated images \cite{yang2024impus, cao2025freemorph}. The variance of the perceptual distance is used in \cite{zhang2024diffmorpher} to  evaluate the regularity of morphing trajectories. In the audio domain, several metrics have been proposed to evaluate audio morphing, related to the concepts of correspondence, smoothness, and intermediateness \cite{niu2024soundmorpherperceptuallyuniformsoundmorphing, kamath2024morphfaderenablingfinegrainedcontrollable}. 
After formally introducing 
those metrics 
we will carefully study their behavior in the remainder of the paper. We believe that they face several limitations in their geometrical design. First, some metrics target only a specific property of a morphing trajectory, such as smoothness, rather than providing a unified measure of whether a sequence of intermediate sounds forms a regular transition between two sounds. Second, we will demonstrate 
that these metrics yield favorable scores for some adversarial morphing trajectories, raising concerns about their robustness. 


We thus introduce a novel objective metric entitled Sobolev Distance to Ideal Morphing (SDIM), that quantifies the regularity of the trajectory between two sounds in an audio embedding space. The metric uses Sobolev distances in multidimensional spaces~\cite{Caamanosobolev2022} to compare the evaluated morphing trajectory with an ideal morphing trajectory. Our contribution is twofold: 1) we first introduce SDIM, a metric to evaluate audio morphing regularity, and 2) we then propose to leverage a physical synthesis model to assess the ability of SDIM to capture morphing regularity, and we compare it to several state-of-the-art morphing metrics. The code and data used to produce the results, as well as sound and morphing trajectory examples are available on the companion page associated to the paper.\footnote{Code, data, and sound examples: \\https://theochaslecauchy.github.io/paperSobolevMorphingMetric}

Those contributions allow us to make significative progresses in directions which are crucial for generative audio.
First, we propose a metric that is robust to subtly adversarial morphing trajectories that are of poor perceptual quality but which still score very high when evaluated using state-of-the-art audio morphing metrics.
Second, we demonstrate with solid empirical evidence that the most recent and deepest neural audio embeddings are more efficient at enabling the proposed SDIM to capture the regularity of morphing trajectories.
The great ability of SDIM metrics to assess jointly the linearity and the uniformity of morphing trajectories will certainly improve as novel finer neural audio representations will be released.

\section{Quality of Morphing Trajectories}

\label{sec:measuring_quality_morphing}

\begin{figure}[t]
    \centering
    \includegraphics[width=\columnwidth]{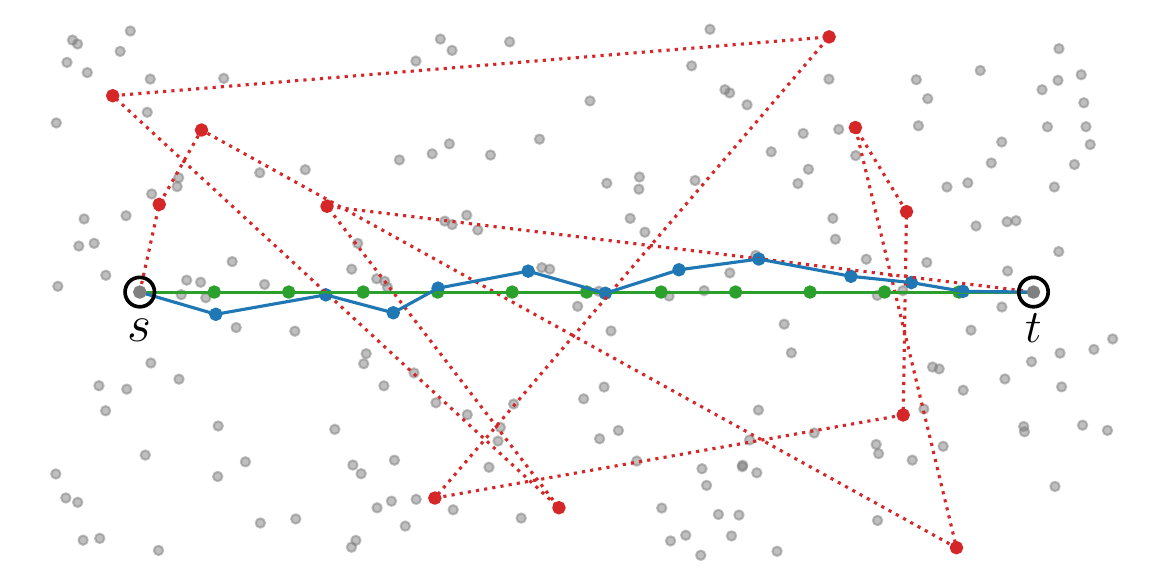}
    \caption{\textbf{Morphing trajectories from two sounds $s$ and $t$ in a perceptually correlated embedding space.}
    Ideal linear regular trajectory (solid green), good quality morphing trajectory (solid blue), and random \Null{} trajectory (dotted red).}
    \label{fig:Morphing}
\end{figure}

The sound morphing task is the synthesis of several audio samples perceived as single-concept sounds which smoothly interpolate between two anchor samples, respectively referred to as the source $s$ and the target $t$~\cite{caetano2013musicalintrumentsoundmorphing}. 
More formally, an audio morphing trajectory is a sequence of $n+2$ audio samples $\{m_0 =s, \hdots, m_{n+1} = t\}$ whose perceptual properties shift gradually from the source to the target. 
In a feature space whose dimensions correlate with human auditory perception, this gradual shift translates into an ideal morphing trajectory consisting in a regular linear interpolation, green dots in Fig.~\ref{fig:Morphing}, from the source $s$ to the target $t$. This ideal morphing trajectory is denoted as Linear and Uniform Morphing (\Refer{}) in the following.
A reference realistic morphing trajectory is expected to deviate only slightly from this ideal path in the perceptual space, as exemplified by the blue dots in Fig.~\ref{fig:Morphing}. Conversely, a very poor morphing trajectory, referred to as \Null{,} can be obtained by sequentially choosing independent random audio samples, red dots in Fig.~\ref{fig:Morphing}, leading to the dotted red trajectory significantly departing from the ideal morphing.

To assess the quality of a morphing trajectory, it is necessary to translate the aforementioned intuition of regular linear interpolation in a perceptual space into precise criteria.
In~\cite{caetano2012formalevaluation}, three properties are proposed to assess the perceptual quality of a morphing trajectory: \Corres{,} \Smooth{,} and \Interm{.}
The \Corres{} evaluates the capacity of the synthesizer to apply the morphing on corresponding parts of the sound, for example a good morphing sound between a violin tone and a flute tone should blend the respective attacks of the violin tone and the flute tone, but should not blend the attack of the violin tone with the sustain part of the flute tone. 
The \Smooth{} assess that the trajectory is perceived as a uniformly regular change between anchors. 
Finally, the \Interm{} evaluates whether the morphing sound is perceived as having intermediate perceptual properties.
The gold standard to compare different morphing techniques would be to perform listening tests in which humans would quantify how much these properties are satisfied. 
Yet, implementing such tests is very costly and prone to large variance between listeners, which motivated the introduction of objective and efficiently computable metrics.



\noindent \textbf{Existing Audio Morphing Metrics.}
Existing audio morphing metrics all follow a common two-stage construction. First, the sounds along a morphing trajectory are mapped to an audio embedding space. Second, a geometric criterion is applied to this trajectory to quantify a specific property of the morphing process. In this formulation, the choice of embedding space and the design of the geometric criterion are largely independent, and will be treated as such throughout the remainder of the paper. A first metric based on \Smooth{}, referred to as \SmoothMF{} in the following, was proposed in \cite{kamath2024morphfaderenablingfinegrainedcontrollable}. \SmoothMF{} computes the distances in the CLAP \cite{elizalde2023claplearningaudioconcepts} embedding space---which has been shown to correlate well with human perception---between the embeddings of the morphed sounds and that of the source sound. It then measures smoothness as the Pearson correlation between these distances and the interpolation parameter. 
Alternatively, the CDPAM loss \cite{manocha2021cdpamcontrastivelearningperceptual} computed in the CDPAM embedding space has been leveraged to evaluate both the \Smooth{} and the \Interm{} properties~\cite{niu2024soundmorpherperceptuallyuniformsoundmorphing}: the \SmoothSM{} metric focuses on the mean and standard deviation of the loss along the trajectory, while the \IntermSM{} metric is defined as the sum of the CDPAM loss along the trajectory.
Finally, the \CorresSM{} metric introduced in~\cite{niu2024soundmorpherperceptuallyuniformsoundmorphing} embeds the trajectory's samples into an Mel-Frequency Cepstral Coefficients (MFCC) space and quantifies how much the midpoint leans toward either the source or the target. A value of 0 indicates that the midpoint is equidistant from the source and target in the embedding space.
It is worth noting that the notion of \Corres{,} as defined perceptually in~\cite{caetano2013musicalintrumentsoundmorphing}, is extremely difficult to model as it amounts to performing a morphological analysis of sounds. In the following, the values of \CorresSM{} will be reported for the sake of completeness keeping in mind that, to the best of our understanding, its mathematical properties do not align well with the expected perceptual properties.



\begin{figure}[t]
    \centering
    \includegraphics[width=0.995\columnwidth]{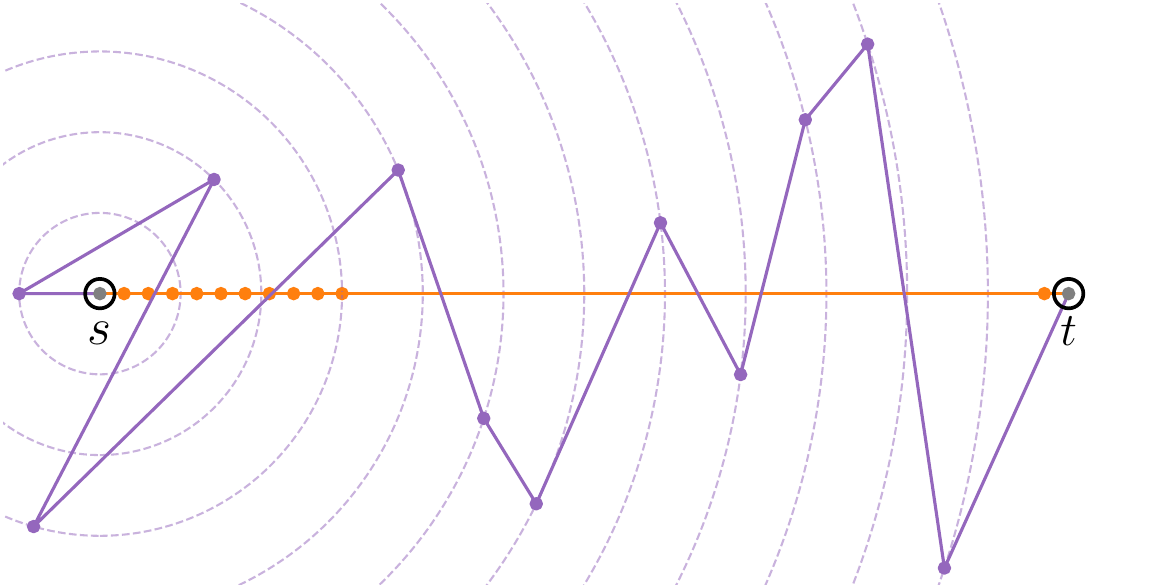}
    \caption{\textbf{Adversarial morphing trajectories in a perceptually correlated embedding space.} (\NUC{,} solid orange) Nonuniform Configuration where the angles are preserved but the spacing regularity is not; (\EQC{,} solid violet) Equidistant Configuration  where the distance to the source is the same as in the ideal trajectory but the angle between $\vec{sm}_i$ and $\vec{st}$ is randomly sampled.}
    \label{fig:AversarialConfigurations}
\end{figure}

\noindent \textbf{Adversarial Trajectories Configurations.}
To assess whether the metrics accurately reflect the ability of the sound synthesizer to generate morphing trajectories that have good \Smooth{} and \Interm{} properties, we design two adversarial configurations of morphing trajectories in the perceptual embedding space, illustrated in Fig.~\ref{fig:AversarialConfigurations}.
Given fixed source $s$ and target $t$,
\begin{itemize}
    \item the Nonuniform Configuration (\NUC{)} consists of a trajectory  in the embedding space in which the $n-1$ first intermediate points are regularly spaced on the segment bridging $s$ and $t$ according to $m_i = s + \frac{0.1}{n}i\times \vv{st}$ for $i\in \llbracket1,n-1\rrbracket$, i.e., very close to $s$, and the last intermediate point is placed at $m_n = s + (1 - \frac{0.1}{n})\times \vv{st}$, i.e., very close to $t$;
    \item in the Equidistant Configuration (\EQC{)} the $i^{\text{th}}$ intermediate point is placed at distance $\frac{i}{n} \times \lVert \vec{st}\rVert$ from the source $s$, as is the $i^{\text{th}}$ point of the ideal trajectory, but the angle between $\vec{st}$ and $\vec{sm}_i$ is sampled randomly, which drastically contrasts with the point alignment in the ideal trajectory.
\end{itemize}

\section{Sobolev Distances to Ideal Morphing}
\label{sec:sobolev}
The overarching goal of this work is to design a morphing metric that jointly measures \Smooth{} and \Interm{} properties, despite them not being directly measurable. 
This require us to rely on the important hypothesis that there exist embedding spaces, whose geometries  correlate with human perception, that capture most of the essence of the \Smooth{} and \Interm{} properties. This hypothesis is largely adopted both in the image and audio processing fields, as it is the basis of widely used metrics such as the Fréchet Inception Distance (FID) \cite{heusel2017gans} or the Fréchet Audio Distance (FAD) \cite{kilgour2019frechet}. 
In Section~\ref{sec:validating_regularity_assumptions}, we provide empirical evidence that several such embeddings satisfy this hypothesis, and we quantify the extent to which they do, enabling their objective comparison.


\noindent \textbf{Sobolev norms.} 
Matching a trajectory requires, first, to follow the same path, e.g., the ideal green line in Fig.~\ref{fig:Morphing}, and second to travel along that path at the same speed, e.g., constant speed for the ideal path in Fig.~\ref{fig:Morphing}, leading to regularly spaced points.
Well adapted tools to assess jointly these two conditions are distances induced by \emph{Sobolev norms}~\cite{Caamanosobolev2022}, which not only take into account functions values but also derivatives up to a given order.
Let $p  \geq 1$, $k \in \mathbb{N}^*$, and $f, g : \Omega \subset\mathbb{R}^d \rightarrow \mathbb{R}^e$ with weak derivatives in $L^p(\Omega)$\footnote{$L^p(\Omega)$ is defined as the space of functions $f :\Omega \rightarrow \mathbb{R}^e$ such that $\lVert f \rVert_{L^p(\Omega)}^p := \int_{\Omega} \lVert f(t)\rVert^p \, \mathrm{d}t < \infty$.} up to order $k \in \mathbb{N}^*$, the Sobolev distance between $f$ and $g$ 
is defined as:
\begin{equation}
    d_{k,p}(f, g) = \sum_{\ell=0}^k \left\lVert D^{(\ell)} f - D^{(\ell)} g \right\rVert_{L^p(\Omega)}
\end{equation}
where $D^{(\ell)}$ denotes the weak derivative operator of order $\ell$, and $\lVert \cdot \rVert_{L^p(\Omega)}$ is the $L^p$ norm on $\Omega$.
For audio morphing, trajectories are parameterized by a unique parameter, thus $d = 1$, and are looked at in an embedding space of dimension $e$.
An original claim of the present work is that the Sobolev distances\footnote{For the sake of simplicity, only the $L^2(\Omega)$ case is considered here, i.e., $p = 2$.} $d_{0,2}$ and $d_{1,2}$ computed in the embedding space provide an \emph{appropriate} measure of similarity between a morphing trajectory to be evaluated and an ideal \Refer{} trajectory. In particular,  small Sobolev distances would indicate that the evaluated morphing trajectory shares the \Smooth{} and \Interm{} properties of the \Refer{} reference.
Compared to $S_{0,2} := d_{0,2}(\cdot, \text{\Refer})$, the SDIM metric $S_{1,2} := d_{1,2}(\cdot, \text{\Refer})$ further penalizes the deviations from the  \Refer{} path, thus quantifying the directness of the morphing trajectory.
In the following, the two proposed metrics $S_{0,2}$ and $S_{1,2}$ are referred to as SDIM metrics.

\section{Physics-based Ideal Audio Morphing}

\label{sec:physics}

The quality of an audio morphing is purely defined from a \emph{perceptual} perspective.
Yet, there exist no inverse transform enabling to go from an ideal trajectory in the embedding space to a set of sounds along the morphing trajectory. This makes automated generation of reference high quality morphing trajectories a challenging task.
To design reference (\Refer{}) trajectories, we propose to leverage an established physical model recognized for its ability to synthesize realistic sounds of plates made from various materials, particularly wood and metal. Each \Refer{} trajectory starts from a wood sound and ends with a metal sound.

This model is formulated as a fourth-order partial differential equation that describes the vibration of a rectangular drum membrane by jointly modeling stiffness, wave propagation speed, first-order damping, and third-order damping. This equation has been solved in~\cite{schaeferphysicalmodelingicsv2019} using the Functional Transfer Method, which is widely used both in 1D~\cite{rabensteinDigitalSoundSynthesis}, 2D~\cite{schaeferphysicalmodelingicsv2019, rabenstein2010spatial}, and 3D~\cite{Rabenstein2010TubularBA} resonant structures sound synthesis. 
Recently, an original reparameterization of this equations has been implemented~\cite{han2024learningtosolvetaslp} involving five interpretable parameters: 1) $\log(\omega)$: the fundamental frequency of the drum, controlling pitch, 2) $\tau$: the duration of the drum sound, inversely proportional to the modulation frequency, 3) $\log(p)$: the frequency-dependent damping coefficient in log-scale, 4) $\log(D)$: the dispersion parameter, controlling the inharmonicity of the overtones in log-scale, and 5) $\alpha$: the aspect ratio of the rectangular drum membrane. As mentioned in~\cite{aramaki2006percussivesoundsynthesizer}, the damping coefficient $\log(p)$ can be used to control the perceived material of the modeled plate. 
By informal listening, we determined that $\log(p) = \log(0.15)$  led to metal-like sounds and $\log(p) = \log(2.0)$ led to wood-like sound. We also defined an admissible range for the remaining parameters: $\log(\omega) \in [\log(1500), \log(3000)]$, $\tau \in [0.2, 0.4]$, $\log(D) \in [-5, -1]$ and $\alpha \in [0.25, 1.0]$. The resulting admissible hypercube is perceptually validated by listening to two sets of sounds $\mathcal{M}$ (metal) and $\mathcal{W}$ (wood), respectively composed of the sixteen sounds resulting from the permutations of the extrema of the ranges of the parameters, except $\log(p)$ which is set to $\log(0.15)$ for $\mathcal{M}$ and $\log(2)$ for $\mathcal{W}$.

\section{Regularity of Neural Audio Embeddings}

\label{sec:validating_regularity_assumptions}

\begin{table}[t]
\begin{center}
{
\begin{tabular}{l|r|r}
Model \hfill ( \# parameters) & \multicolumn{1}{c|}{$S_{0,2}^{\text{\Refer}}/S_{0,2}^{\text{\Null}} \downarrow$} & \multicolumn{1}{c}{$S_{1,2}^{\text{\Refer}}/S_{1,2}^{\text{\Null}} \downarrow$} \\ \hline
MFCC \cite{daviscomparisonofparametricrepresentations1980} \hfill (-) & .18 (.07) & .07 (.03) \\ \hline
VGGish \cite{hershey2017cnnarchitectureslargescaleaudio}  \hfill (72M) & .23 (.08) & .11 (.04)\\ \hline
MSCLAP \cite{elizalde2023claplearningaudioconcepts}  \hfill (159M) & .15 (.03) & .08 (.01)\\
CLAP-Music \cite{wulargescalecontrastive2023}  \hfill (158M) & .23 (.05) & .15 (.03) \\
CLAP-Audio \cite{wulargescalecontrastive2023} \hfill  (158M) & .19 (.06) & .12 (.03) \\ \hline
MERT-v0-public \cite{li2024mertacousticmusicunderstanding}  \hfill (95M) & .12 (.05) & .04 (.02) \\
MERT-v1-95M \cite{li2024mertacousticmusicunderstanding}  \hfill (95M) & .12 (.05) & .05 (.02) \\
MERT-v1-330M \cite{li2024mertacousticmusicunderstanding} \hfill  (330M) & \textbf{.06 (.02)} & \textbf{.02 (.01)} \\
\end{tabular}
}
\caption{\textbf{Normalized SDIM metrics for different embeddings.}
Normalized SDIM metrics means (standard deviations) are estimated from $1,000$ \Refer{} trajectories and $1,000$ \Null{} trajectories.
}
\label{tab:embeddings_linearity}
\end{center}
\end{table}



The choice of embedding is critical, as shown by the literature on metrics such as the FAD \cite{gui2024adapting, tailleur:hal-04603443}. Here, the ability of SDIM metrics to assess morphing quality is tightly bound to the ability of the chosen embedding to capture the fine perceptual properties that should evolve linearly and uniformly along good morphing trajectories (e.g. from wood-like to metal-like sounds). 
An embedding that correlates well with subtle perceptual cues should thus lead to ideal morphing trajectories achieving small $S_{0,2}$ and $S_{1,2}$ scores. 

The present section proposes a systematic comparison of several popular audio embeddings through the prism of the regularity of ideal morphing trajectories.
For the sake of completeness, audio embedding spaces covering both general-purpose and music-specific representations are considered.
As general-purpose representations, we consider Mel-Frequency Cepstral Coefficients (MFCC)~\cite{daviscomparisonofparametricrepresentations1980}, VGGish \cite{simonyan2015deepconvolutionalnetworkslargescale}, and Microsoft CLAP (MS-CLAP) \cite{elizalde2023claplearningaudioconcepts}. VGGish was trained on a subset of YouTube-100M~\cite{hershey2017cnn}, comprising approximately $350,000$ hours of audio, while MS-CLAP was trained using $4.6$M paired audio-text data. We additionally investigate LAION CLAP \cite{wulargescalecontrastive2023}, using its ``$630$k-audioset-best" model, which was trained on approximately $10,000$ hours of general audio. We also include the music-specific LAION CLAP model ``music-audioset-epoch15-esc-9.14", which was fine-tuned on music data. For music-specific representations, we further consider MERT \cite{li2024mertacousticmusicunderstanding}. MERT provides models with different scales and training data, allowing us to investigate the influence of both factors on the regularity of the resulting embedding space. We evaluate the three models released with the original work: MERT-v0-public, MERT-v1-95M, and MERT-v1-330M. MERT-v0-public was trained on approximately $910$ hours of music from the publicly available Music4All dataset \cite{santanammusic4all}. MERT-v1-95M was trained on approximately $1,000$ hours of music from a larger in-house dataset. Finally, MERT-v1-330M is the largest model, and was trained on approximately $160,000$ hours of privately collected music data.

To compare the regularity of ideal trajectories in the different perceptual spaces induced by these representations, we propose to consider morphing trajectories evolving continuously from a metal-like to a woody-like sound in a perceptual space. To that aim, $1,000$ couples of audio anchors $(s,t)$, with $s$ (resp. $t$) a metal(resp. wood)-like sound, are defined by specifying for each the parameters of the synthesizer of Sec.~\ref{sec:physics} that are used to generate them.  
For each couple, the source $s$ and target $t$ share the same $\log(\omega)$, $\tau$, $\log(D)$ and $\alpha$ design parameters, sampled uniformly in the admissible hypercube, and $\log(p) = \log(0.15)$ (resp. $\log(p) = \log(2)$) for the source (resp. the target).
Audio morphing trajectories are then computed by regularly varying $\log(p)$ while keeping $\log(\omega)$, $\tau$, $\log(D)$ and $\alpha$ to the values used to generate $s$ and $t$.
To ensure a fair comparison of representations which have very different dimensionalities, the proposed SDIM metrics of the \Refer{} trajectories reported in Tab.~\ref{tab:embeddings_linearity} are normalized by their average value on purely random \Null{} trajectories $S_{k,2}^{\text{\Null}}$.
Both the $0^{\text{th}}$-order $S_{0,2}$ and $1^{\text{st}}$-order $S_{1,2}$ SDIM metrics consistently yield low values for MERT neural representations, with their respective lowest values obtained for the largest MERT-v1-330M~\cite{li2024mertacousticmusicunderstanding}.
This indicates their ability to accurately capture the smooth evolution along the audio morphing trajectory. On the contrary, the hand-crafted MFCC representation and the light neural embedding VGGish lead to large normalized SDIM values. The more recent CLAP and MS-CLAP embeddings, which also predate MERT and are relatively large, similarly lead to large normalized SDIM values. Using these embeddings with SDIM would thus erroneously characterize \Refer{} trajectories as poor morphings. As MERT-v1-330M captures the best the regularity of \Refer{} trajectories, it is used in all following experiments.


\section{Sensitivity to trajectories configurations in the audio domain}

\label{sec:sensitivity_to_trajectories}

\begin{table}[t]
\centering
\resizebox{\columnwidth}{!}
{
    \begin{tabular}{l|r|r|r|r}
        \multicolumn{1}{c|}{Metric} & \multicolumn{1}{c|}{\Refer} & \multicolumn{1}{c|}{\NUC{}} & \multicolumn{1}{c|}{\EQC{}} & \multicolumn{1}{c}{\Null{}}\\ \hline
        \corres{} $\downarrow$ & .19 (.02) & .38 (.01) & \textcolor{red}{\underline{.14 (.03)}} & \textcolor{red}{\underline{.03 (0.02)}} \\
        \smoothMF{} $\uparrow$ & .96 (.01) & .75 (.01) & \textcolor{red}{\underline{.97 (.03)}} & .11 (0.17)\\
        \interm{} $\downarrow$ & .23 (.04) & \textcolor{red}{\underline{.21 (.04)}} & 1.15 (.24) & 5.59 (0.73) \\
        \smoothSM{} $\downarrow$ & .02 (.00) & \textcolor{red}{\underline{.02 (.00)}} & .09 (.02) & .52 (0.07) \\ \hline
        $10^{-2} \times S_{0,2}$ $\downarrow$ & .69 (.09) & 1.14 (.20) & 1.72 (.32) & 8.41 (1.27) \\
        $10^{-3} \times S_{1,2}$ $\downarrow$ & .35 (.05) & .69 (.12) & 2.59 (.51) & .013 (1.54) \\
    \end{tabular}
}
\caption{\textbf{Average metrics on the four configurations of trajectories in the MERT embeddings space.} Means (standard deviations) estimated from $1000$ examples. Only the SDIM metrics (last two rows) are robust to all of the adversarial configurations. Underlined red font indicates failure in measuring the true regularity.}
\label{tab:audio_configurations}
\end{table}

Once chosen an appropriate perceptual audio embedding, it is possible to compare the proposed SDIM metrics to state-of-the-art ones on audio morphing trajectories.
We evaluate the ability of the different metrics to discriminate between ideal  linear and uniform, \Refer{} trajectories and nonuniform \NUC{}, nonlinear \EQC{}, and purely random \Null{} trajectories.
The \Refer{} and \Null{} trajectories are generated following the procedure described in Sec.~\ref{sec:validating_regularity_assumptions}. The design of \NUC{} trajectories is controlled in the synthesizer's parameter space. The parameter $\log(p)$ is varied nonuniformly following the principle described at the end of Sec.~\ref{sec:measuring_quality_morphing}. Sounds are then generated for each parameter value, and their embeddings are computed to form a trajectory in the embedding space.
As for the \EQC{} trajectories, they are selected a posteriori in the embedding space from a large set of $50,000$ sounds synthesized using the physics-based generator of Sec.~\ref{sec:physics} with parameters randomly sampled within the admissible hypercube.
The $i^{\text{th}}$ point $m_i$ of an \EQC{} trajectory is chosen so that the distance $\lVert \vec{sm}_i\rVert$, computed in the embedding space,  matches as closely as possible the distance between the source and the $i^{\text{th}}$ point of the ideal \Refer{} trajectory.
This strategy enables to implement approximately the \EQC{} setup described in~Sec.~\ref{sec:measuring_quality_morphing}.

\begin{figure}[t]
    \centering
    \includegraphics[width=\columnwidth]{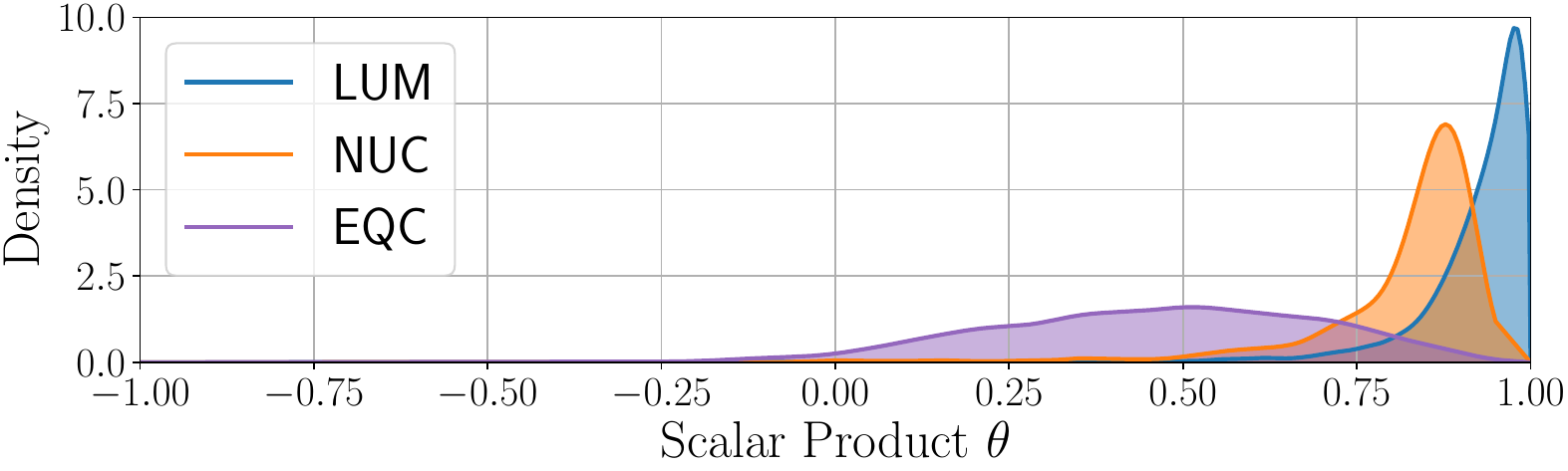}
    \caption{\textbf{Angle dispersion for the \Refer{}, \EQC{},  and \NUC{} configuration trajectories.}
    Angles are measured through scalar products between the segment connecting the source and the intermediate point and the source and the target (ideal linear trajectory). Histograms are computed over $11$ intermediate points across $1000$ trajectories.}
    \label{fig:scalar_products_densities}
\end{figure}

These two sampling procedures are validated by inspecting the angle, computed in the embedding space, between the line connecting the source $s$ and the target $t$ and that connecting the source and the current point $m_i$ of the trajectory through:
\begin{equation}
\theta(i)=\left\langle\frac{\vec{sm}_i}{||\vec{sm}_i||}, \frac{\vec{st}}{||\vec{st}||}\right\rangle
\end{equation}
where $\theta=1$ (resp. $\theta=-1$)  indicates perfect (resp. anti-)alignment between $\vec{st}$ and $\vec{sm}_i$. Histograms of these angular indicators are displayed in Fig.~\ref{fig:scalar_products_densities}. The reference trajectory has an angular indicator highly concentrated around $1$, indicating that it is almost exactly linear. The angular indicator for \NUC{} is also close to $1$, although slightly less concentrated, suggesting that these trajectories are satisfactorily linear. Finally, the \EQC{} trajectories have highly dispersed angular indicators, showing that they are clearly not linear.



The six metrics averaged over $1,000$ realizations of each trajectory type are reported in Tab.~\ref{tab:audio_configurations}.  Only the SDIM metrics are robust to all of the adversarial configurations. While not reported in the paper for conciseness, equivalent results were obtained when computing the state-of-the-art metrics in their original embedding spaces.




\section{Conclusion}

In this work, we introduced Sobolev Distances to Ideal Morphing (SDIM), objective metrics for evaluating the regularity of audio morphing trajectories. The key asset of SDIM metrics is  their theoretically grounded geometrical properties that allow us to jointly measure linearity and uniformity properties, leading to a one figure metric, ideal for ranking alternative designs of sound synthesizers.


Through experiments based on a physics-based ideal audio morphing model, we demonstrated that SDIM metrics computed in an appropriate space can discriminate between morphing trajectories with different degrees of regularity. In particular, SDIM metrics remain robust to specific adversarial trajectories contrary to state-of-the-art audio morphing metrics, highlighting the importance of explicitly considering the overall trajectory when evaluating morphing regularity. We further show that recent and larger audio embedding models tend to produce higher regularity scores on controlled regular morphing trajectories, suggesting that advances in audio representation learning may provide increasingly suitable spaces for objectively evaluating morphing regularity with SDIM metrics.

Future work will investigate the relationship between SDIM metrics and human perceptual judgments of morphing regularity. We further hypothesize that, leveraging the generality of the Sobolev norms, SDIM metrics can be applied to any regular and perceptually informed embedding space to evaluate morphing regularity, including those used for images and videos.

\clearpage
\newpage


\label{sec:refs}
\bibliographystyle{myIEEEbib}
\bibliography{strings,refs}

\end{document}